\documentstyle[11pt,newpasp,twoside,epsf]{article}
\def\edcomment#1{\iffalse\marginpar{\raggedright\sl#1\/}\else\relax\fi}
\reversemarginpar

\long\def\symbolfootnotemark[#1]{\begingroup%
\def\thefootnote{\fnsymbol{footnote}}\footnotemark[#1]\endgroup}
\long\def\symbolfootnotetext[#1]#2{\begingroup%
\def\thefootnote{\fnsymbol{footnote}}\footnotetext[#1]{#2}\endgroup}

\begin{document}
\title{Relativistic Outflows in AGNs}
\author{Nektarios Vlahakis\symbolfootnotemark[2] \& Arieh K\"onigl}
\affil{Department of Astronomy \& Astrophysics, 
University of Chicago, 5640 S. Ellis Ave., Chicago, IL 60637, USA}
\symbolfootnotetext[2]{Present address: Section of Astrophysics, Astronomy \& Mechanics, 
Department of Physics, University of Athens, 15784 Zografos Athens, Greece}

\begin{abstract}
There are observational indications that relativistic outflows in AGNs are
accelerated over distances that far exceed the scale of the central
engine. Examples include the radio galaxy NGC 6251, where knots in the
radio jets were inferred to accelerate from $\sim 0.13\ $c at a distance of
$\sim 0.53\ {\rm pc}$ from the galactic nucleus to $\sim 0.42\ $c at $r=1.0\ {\rm pc}$,      
and the quasar 3C 345, 
where the Lorentz factor of the radio knot C7 was deduced to increase from
$\sim 5$ to $>10$ as it moved from $r=3\ {\rm pc}$ to $r=20\ {\rm pc}$.
It is argued, using exact semianalytic solutions of the relativistic MHD equations, that this
behavior is a signature of magnetic acceleration. The same basic
driving mechanism may apply to the relativistic jets in AGNs, gamma-ray burst sources, and
microquasars.
\end{abstract}

\section{Introduction}
Magnetic acceleration has long been
considered an attractive mechanism for the origin of AGN jets
(e.g., Blandford \& Payne 1982). These models have commonly been
thought to imply that the bulk of the acceleration occurs fairly
close to the source, as in thermally driven outflows.
Since the region near the central
black hole cannot be resolved, it has been unclear whether one could
observationally infer that magnetic driving actually takes place.
Recent theoretical work 
has, however, established that magnetic acceleration can occur
over many decades in radius, for both nonrelativistic (Vlahakis et al. 2000) 
and relativistic (Li et al. 1992; 
Vlahakis \& K\"onigl 2003a, hereafter VK) outflows.
There are already a number of AGNs for which there is 
direct observational evidence for acceleration to relativistic speeds 
occurring in an extended (parsec-scale) region. 
The acceleration on scales that are much larger than the gravitational 
radius of the central object cannot be easily explained in terms of thermal 
driving (Vlahakis \& K\"onigl 2003b). Thus, magnetic driving is
the most plausible interpretation of the observed 
parsec-scale accelerations. Here we discuss two examples (a radio
galaxy and a blazar) and argue that they provide strong support
for magnetic stresses as the dominant acceleration mechanism.

\section{The Relativistic MHD Model and Applications to AGN Outflows}
The system of equations of special-relativistic,
ideal MHD consists of the Maxwell and Euler equations together with
the mass and specific-entropy conservation relations.
Assuming axisymmetry [$\partial / \partial \phi=0$
in spherical $(r\,,\theta\,,\phi)$ and cylindrical
$(\varpi\,,\phi\,, z)$ coordinates]
and a steady state,  the full set of equations can
be partially integrated to yield several field-line constants.
These constants are the total specific angular momentum $L(A)$,
the field angular velocity $\Omega(A)$,
the ``magnetization parameter'' $\sigma_{\rm M}(A)$
(with the mass-to-magnetic flux ratio given by ${A\Omega^2}/{\sigma_{\rm M} c^3}$),
the adiabat $Q(A)\equiv P/ \rho_0^{5/3}$,
and the total energy-to-mass flux ratio $\mu(A) c^2
=\xi \gamma c^2 + ({c}/{4 \pi}) ({E |B_\phi|}/{\gamma \rho_0 V_p})$.
Here $A$ is the poloidal magnetic flux function (which
identifies the field line), $\xi c^2=  c^2+ {5 P}/{2\rho_0}$ is the specific enthalpy,
${\bf {V}}$ is the bulk velocity, $\gamma$ is the Lorentz factor, 
$\rho_0$ and $P$ are the comoving matter density and pressure,
${\bf {E}}$ and ${\bf {B}}$ are the electric and magnetic fields,
and the subscripts $p$ and $\phi$ denote poloidal and azimuthal
vector components, respectively. VK integrated the two remaining relations
(the Bernoulli and transfield force-balance equations)
assuming radial self-similarity of the form $r={\cal F}_1(A) {\cal F}_2 (\theta)$.
With this ansatz it is possible to separate the $(A\,, \theta)$ variables
if the following relations hold: ${\cal F}_1(A) \propto A^{1/F}$,
$L(A) \propto A^{1/F}$, $\Omega(A) \propto A^{-1/F}$, $Q(A) \propto A^{-2(F-2)/3}$,
$\mu(A)= const$, and $\sigma_{\rm M}(A)  = const$
(see Li et al. 1992 and Contopoulos 1994 for the ``cold'' limit of this model).
The parameter $F$ controls the current distribution:
$\varpi |B_\phi| = A^{1-1/F} {\cal F}(\theta)$.
Close to the origin the field is force-free: ${\cal F}(\theta)
\approx const$, and hence $\varpi |B_\phi| \propto A^{1-1/F}$.
Thus, the parameter regime $F>1$ corresponds to a
current-carrying jet, which is applicable near the rotation
axis, whereas $F<1$ corresponds to the return-current regime, which
possibly occurs at large cylindrical distances.
The remaining task is to integrate ordinary differential equations
and find solutions crossing the Alfv\'en and modified-fast magnetosonic singular points.
We now apply this formalism to the interpretation of
accelerating relativistic jets in two AGNs, the 
powerful radio galaxy NGC 6251 and the quasar 3C 345.

VLBI measurements indicate sub--parsec-scale acceleration of the
radio jet and counterjet in NGC 6251 from
{$V(r=0.53\ {\rm pc}$$)=0.13 c$} to
{$V(r=1\ {\rm pc}$$)=0.42 c$} (Sudou et
al. 2000).
Magnetic acceleration can naturally account for these
observations, as demonstrated with a specific solution in Figure
1.  Panels ($c$)--($g$) in this figure show various quantities
as functions of $\varpi/\varpi_{\rm A}$
(which, in turn, is a function of the polar angle $\theta$)
along the outermost field line. (Here $\varpi_{\rm A}$ is the Alfv\'en lever arm, and
$\varpi_{\rm A, out} = 10^3\, \varpi_{\rm A, in} = 0.98\ {\rm
pc}$ in this example.) Panel ($c$) depicts the force densities in the poloidal direction,
showing that thermal and centrifugal effects are important only
near the origin, with the magnetic pressure-gradient force rapidly
becoming the dominant driving mechanism. The same conclusion can
be drawn from the plot of the energy flux components in panel
($d$): acceleration driven by the conversion of enthalpy into kinetic energy
occurs only near the origin, where $\xi \gamma > \gamma$.
In this region the field is nearly force-free and the
Poynting-to-mass flux ratio remains roughly constant. Beyond the
point where $\xi \gamma \approx \gamma$
the magnetic acceleration takes over: in this regime the increase in the
Lorentz factor results from a decrease in the Poynting-to-mass flux ratio.
Asymptotically, an approximate equipartition between the kinetic
and Poynting fluxes is attained.  Panels ($e$) and ($f$) depict the bulk velocity
components and the temperature, respectively. Even if the initial temperature is as
high as $\sim 10^{12}\ {\rm K}$, thermal effects are overall
insignificant. Panel ($g$) shows that the magnetic field
is primarily poloidal near the origin of the flow but
becomes predominantly azimuthal further downstream.
Asymptotically, $B_z \propto \varpi^{-2}\,, -B_\phi \propto
\varpi^{-1}$; also $B_\varpi << B_z$ ---  a signature of cylindrical collimation.
\begin{figure}[h]
\plotone{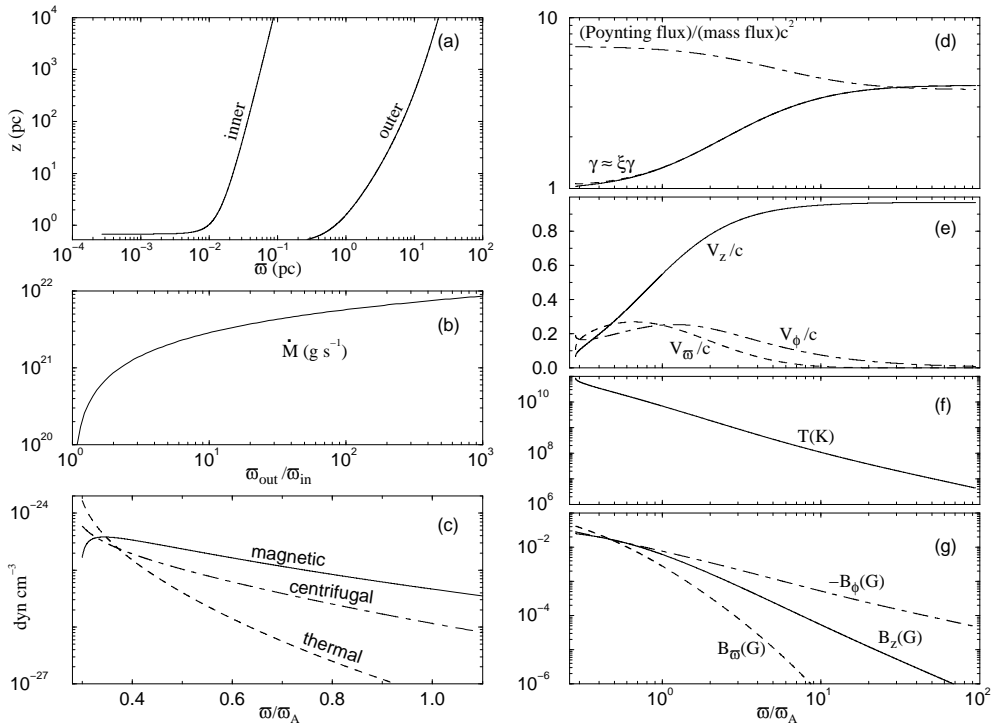}
\caption{$r$ self-similar solution describing the jets in NGC 6251.
$(a)$ Poloidal field-line shape on a logarithmic scale.
$(b)$ Mass-loss rate as a function of ${\varpi_{\rm out}}/{\varpi_{\rm in}}$, the ratio of the
outermost and innermost disk radii. The remaining panels are
discussed in the text.}
\end{figure}

VLBI images of the quasar 3C 345 indicate that 
the jet components' speeds increase with separation from the core
(Zensus, Cohen, \& Unwin 1995; Lobanov \& Zensus 1999). In particular, in
the case of the C7 component, Unwin et al. (1997) inferred that it
accelerates from $\gamma \sim 5$ to $\gamma \ga 10$
over the (deprojected) distance range (measured from the core)
$\sim 3-20\ {\rm pc}$.
We propose that this pc-scale acceleration is again most
plausibly interpreted in terms of
magnetic driving, and we present in Figure 2 a particular solution for a
magnetically driven proton-electron jet that supports this conclusion.
For the adopted fiducial parameters, the C7 component is
expected to continue to accelerate up to $\gamma_\infty \approx 30$:
values of this order may, in fact, characterize the more distant
components (in particular, C3 and C5) of the 3C 345 jet (Lobanov \& Zensus 1999).
The pc-scale helical field morphology implied by our model is consistent with
VLBI polarization maps of BL Lac objects (e.g., Gabuzda,
Pushkarev, \& Cawthorne 2000) and circular polarization measurements of blazars
(e.g., Homan, Attridge, \& Wardle 2001). As discussed in VK, the
same basic model may also account for the collimated relativistic
outflows in gamma-ray burst sources and microquasars.
\begin{figure}[h]
\plotone{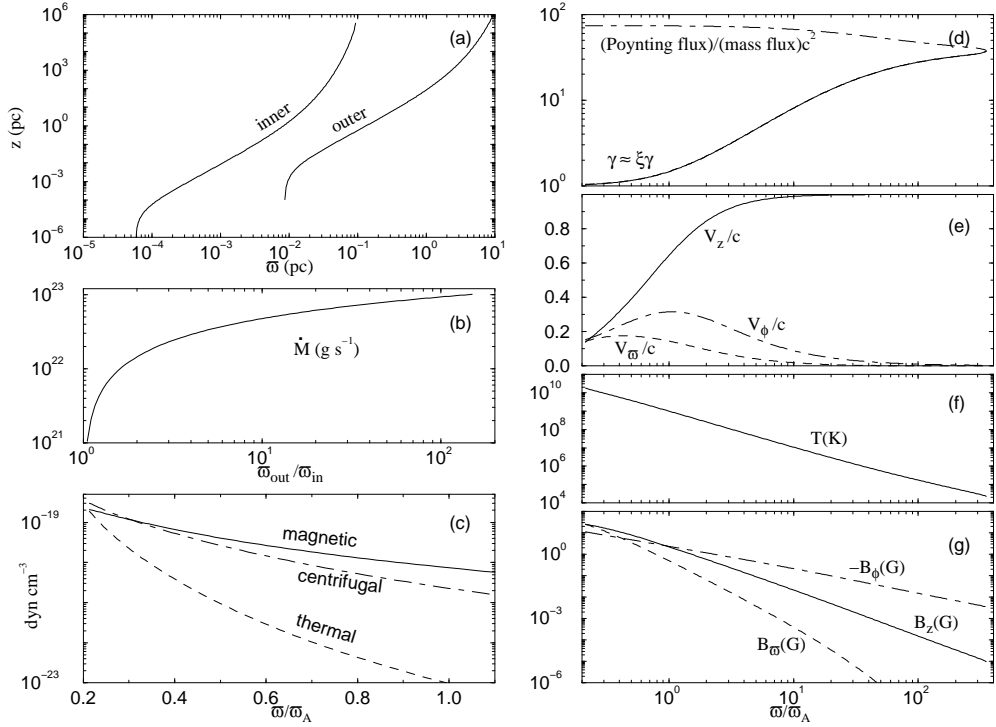}
\caption{
Same as Fig. 1, but for the application to
the superluminal jet in 3C 345. Here $\varpi_{\rm A, out} =
150\, \varpi_{\rm A, in} = 4.1 \times 10^{-2}\ {\rm pc}$.
}
\end{figure}

\end{document}